\documentclass[aps,pra,showpacs,twocolumn]{revtex4-1}
\usepackage{amsfonts}
\usepackage{amssymb}
\usepackage{amsmath}
\usepackage{graphicx}
\usepackage{epsfig}
\usepackage{color}
\usepackage{amsmath,bm}
\usepackage{booktabs}
\usepackage[colorlinks=true, linkcolor=blue, citecolor=blue, urlcolor=blue]{hyperref}

\begin{document}

\title{Condensate ground states of hardcore bosons induced by an array of
impurities}
\author{J. Y. Liu-Sun}
\author{E. S. Ma}
\author{Z. Song}
\email{songtc@nankai.edu.cn}

\begin{abstract}
Neither hardcore bosons nor fermions can occupy the same lattice site-state.
However, a nearest-neighbour interaction may counteract the hardcore effect,
resulting in condensate states in a bosonic system. In this work, we unveil
the underlying mechanism by developing a general method to construct the
condensate eigenstates from those of sub-Hamiltonians. As an application, we
find that a local on-site potential can induce an evanescent condensate
mode. Based on this, exact condensate ground states of hardcore bosons,
possessing off-diagonal long-range order, can be constructed when an array
of impurities is applied. The effect of the off-resonance impurity on the
condensate ground states is also investigated using numerical simulations of
the dynamic response.
\end{abstract}

\affiliation{School of Physics, Nankai University, Tianjin 300071, China}

\maketitle

\section{Introduction}

Bose-Einstein condensation (BEC) serves as a striking example of quantum
phenomena that become evident on a macroscopic scale, as first demonstrated
by Bose and Einstein in their seminal works \cite{bose1924plancks,einstein1924quantentheorie}. 
This extraordinary state of matter highlights the profound impact of quantum
mechanics on the behavior of particles at a level observable to the naked
eye. Specifically, BEC is marked by the formation of a coherent quantum
state among a collection of free bosons, resulting in a remarkable
synchronization of their behavior. Significantly, advancements in cold atom
experiments have greatly propelled the theoretical study of BEC, offering a
highly adaptable framework for creating diverse phases of both interacting
and non-interacting bosonic systems \cite{bloch2012quantum,atala2014observation,aidelsburger2015measuring,stuhl2015visualizing}. Advances in cooling and trapping atoms and molecules with dipolar
electric or magnetic moments enable the realization of extended Hubbard
models featuring density-density interactions 
\cite{goral2002quantum,moses2015creation,moses2017new,baier2016extended,reichsollner2017quantum,Chomaz_2023}. 
Moreover, contemporary experimental setups enable precise control over both
the geometry and interactions, allowing for the direct investigation of the
real-time evolution of quantum many-body systems using engineered model
Hamiltonians \cite{jane2003simulation,bloch2012quantum,blatt2012quantum}. 
In this scenario, a boson within the optical lattice essentially corresponds
to a cluster comprising an even number of fermions. This should lead to
on-site repulsive interactions within the framework of the tight-binding
description, causing an atom to become a hardcore boson in the strong
interaction limit. Most theoretical studies concentrate on the phase
diagram of the ground state over the past several decades 
\cite{J.K.Freericks_1994,batrouni1995supersolids,frey1997critical,kuhner1998phases,batrouni2000phase,hebert2001quantum,D.L.Kovrizhin_2005,schmidt2006single,chen2008supersolidity,capogrosso2010quantum,jiang2012pair}. 
Intuitively, one might expect that on-site repulsive interactions would
prevent the formation of BEC at moderate particle densities, because neither
hardcore bosons nor fermions can occupy the same lattice site-state.
However, it has been shown that a nearest-neighbour (NN) interaction may
counteract the hardcore effect, resulting in condensate states with
off-diagonal long-range order (ODLRO) in a bosonic system \cite{zhang2025exacteigenstatesoffdiagonallongrange}. 
There are two key restrictions on these findings. 

\begin{figure}[t]
\centering
\includegraphics[width=0.45\textwidth]{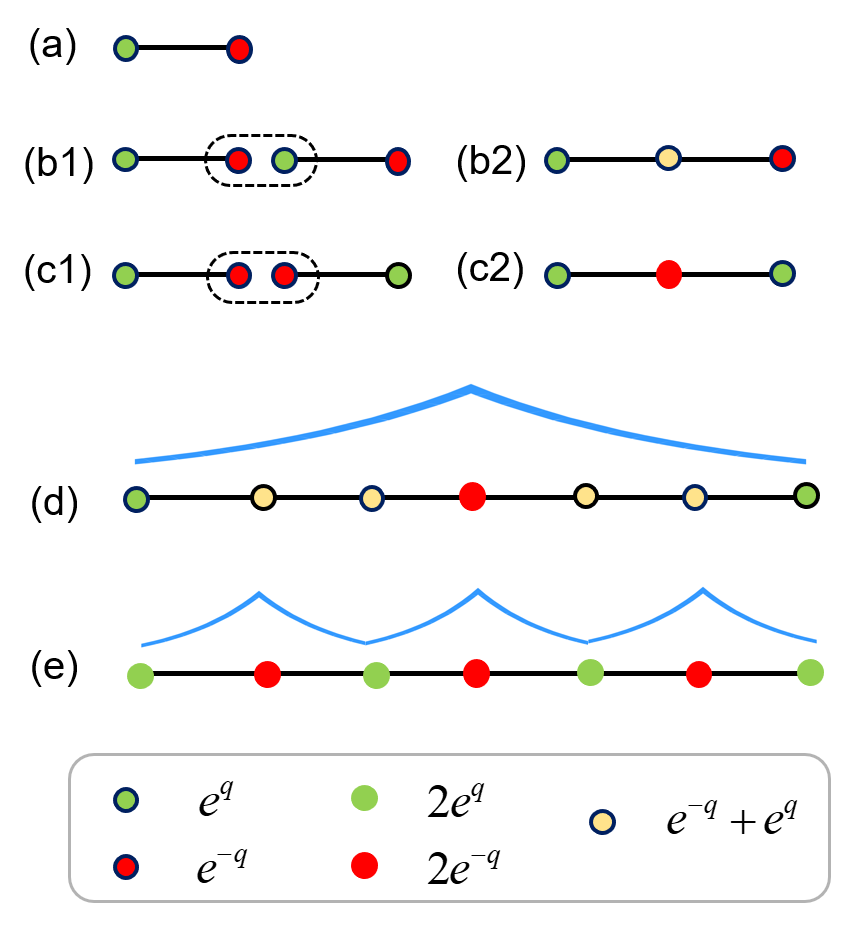}
\caption{Schematic illustrations for constructing a Bose Hamiltonian by a
set of sub-Hamiltonians of dimers. Here, only the on-site potentials are
indicated by the color circles at the bottom. All the schematics are
applicable both for the free-boson system in Section II and the
hardcore-boson system with resonant nearest-neighbor interactions in Section
III. (a) A 2-site system as a basic building block. (b1) The summation of
two dimers by combining the two sites enclosed by the dotted loop, resulting
in a trimer (b2). (b1) and (b2) represent the other way of the summation.
(c) The summation of multi dimers, resulting in a chain with impurities at
the center and two ends. The profile of single-particle ground state is
indicated by the blue tent curve. (d) Alternative combination of multi
dimers, resulting in a chain with impurity array. The corresponding
single-particle ground state is an extended state.}
\label{fig1}
\end{figure}
First, the strength of the
nearest-neighbor (NN) interaction must match the single-particle dispersion
relation. Second, the obtained eigenstates correspond to excited states,
rather than ground states. In addition, rigorous results for a model
Hamiltonian play an important role in physics and sometimes open new avenues
for exploration in the field. The exact solution to the quantum harmonic
oscillator has played a crucial role in the history of physics. It stands as
a key concept in traditional quantum mechanics and continues to be a
cornerstone for modern research and applications within the field.

In this work, we explore the influence of impurities on the formation of
condensate states in one-dimensional systems with NN interaction beyond the
resonant region. To this end, we develop a general method to construct the
condensate eigenstates from those of sub-Hamiltonians. This method unveils
the underlying mechanism for the obtained examples of exact condensate
eigenstates, such as the $\eta $-pairing eigenstates in the Hubbard model 
\cite{Yang1989}. 
We apply the method to the hardcore-boson model with NN interaction beyond
the resonant region. We find that a local on-site potential can induce an
evanescent condensate mode. Based on this, exact condensate ground states of
hardcore bosons, possessing off-diagonal long-range order, can be
constructed when an array of impurities is applied. The effect of the
off-resonance strength of the on-site potentials on the condensate ground
states is also investigated using numerical simulations of the dynamic
response.
This paper is organized as follows. In Sec. \ref{General formalism}, we
present a theorem and demonstrate it for discrete quantum systems. Based on
this, the condensate eigenstates of a Hamiltonian can be constructed from
those of sub-Hamiltonians. In Sec. \ref{Hardcore boson systems}, we apply
the theorem to hardcore-boson systems. In Sec. \ref{Condensate ground states
with ODLRO}, we present the main results of this work: the condensate ground
states with ODLRO. In Sec. \ref{Dynamic stability}, we conduct numerical
simulations to investigate the dynamic stability of the condensate states
when the system is off-resonance. Finally, we present a summary of our
results in Sec. \ref{Summary}.

\section{General formalism}

\label{General formalism}

While exact solutions for quantum many-body systems are uncommon, they play
a crucial role in offering valuable insights into the characterization of
novel quantum matter and its dynamic behaviors. For instance, the exact $%
\eta $-pairing eigenstates of a Hubbard model are a paradigm to demonstrate
Fermi condensation with ODLRO \cite{Yang1989}. 
Additional examples are provided in recent works \cite{ma2022steady,zhang2025exacteigenstatesoffdiagonallongrange,Ma_2024,ma2023off,ma2023polarity,zhang2025coalescing}. 
In this section, we aim to elucidate the common features among these
examples.\ We begin with a general result for the eigenstates of discrete
quantum systems. Based on this result, the eigenstates of a Hamiltonian can
be constructed from those of sub-Hamiltonians. This approach is applicable
to more generalized fermion and boson systems, with no specific restrictions
on dimensionality or geometry.

\textit{Theorem}. Considering a Hamiltonian on a set of lattice sites $%
(a,b,c)$, consisting two sub-Hamiltonians, given by
\begin{equation}
H(a,b,c)=H_{1}(a,c)+H_{2}(b,c),
\end{equation}%
where $a$, $b$, and $c$ label three sub-lattices. Suppose that each
sub-Hamiltonian has a set of $\left( N+1\right) $-fold degenerate
zero-energy eigenstates in the ladder form, that is%
\begin{eqnarray}
H_{1}(s_{a}+s_{c})^{m}\left\vert G\right\rangle &=&0,  \notag \\
H_{2}(s_{b}+s_{c})^{m}\left\vert G\right\rangle &=&0,
\end{eqnarray}%
with $m\in \left[ 0,N\right] $, where $s_{\alpha }$\ ($\alpha =a,b,c$) is a
local operator on the sub-lattices $\alpha $, obeying%
\begin{eqnarray}
\left[ s_{a},s_{c}\right] &=&\left[ s_{a},s_{b}\right] =\left[ s_{b},s_{c}%
\right] =0,  \notag \\
\lbrack H_{1},s_{b}] &=&[H_{2},s_{a}]=0,  \label{identities}
\end{eqnarray}%
and $\left\vert G\right\rangle $\ is a common eigenstates of $H_{1}$ and $%
H_{2}$ on lattice sites $(a,b,c)$. Then there exists a set of zero-energy
eigenstates of $H$, which are constructed through the operator $%
s_{a}+s_{b}+s_{c}$, given by%
\begin{equation}
H(a,b,c)(s_{a}+s_{b}+s_{c})^{m}\left\vert G\right\rangle =0.
\end{equation}%
The proof of this theorem is straightforward. In fact, based on the above
identies, given by Eq. (\ref{identities}), we have 
\begin{eqnarray}
&&H(s_{a}+s_{b}+s_{c})^{m}\left\vert G\right\rangle  \notag \\
&=&H_{1}\sum_{k=0}^{m}C_{m}^{k}s_{b}^{m-k}(s_{a}+s_{c})^{k}\left\vert
G\right\rangle +  \notag \\
&&H_{2}\sum_{k=0}^{m}C_{m}^{k}s_{a}^{m-k}(s_{b}+s_{c})^{k}\left\vert
G\right\rangle ,
\end{eqnarray}%
which equals zero.

This conclusion has following implications. First, assuming that the
groundstate energies of $H_{1}$ and $H_{2}$ are $E_{1g}$ and $E_{2g}$,
respectively, for any given state $\left\vert \phi \right\rangle $, we have
\begin{equation}
\left\langle \phi \right\vert \left( H_{1}+H_{2}\right) \left\vert \phi
\right\rangle =\left\langle \phi \right\vert H_{1}\left\vert \phi
\right\rangle +\left\langle \phi \right\vert H_{2}\left\vert \phi
\right\rangle \geqslant E_{1g}+E_{2g}.
\end{equation}%
In the case that the equality holds, the state $\left\vert \phi
\right\rangle $\ must be the ground state of $H_{1}+H_{2}$. Second, we note
that the explicit form of the operators $\left\{ s_{a},s_{b},s_{c}\right\} $
is unrestricted, which allows our conclusion to be generalized to
high-dimensional systems. Third, the upper bound of $m$\ for the state $%
(s_{a}+s_{b}+s_{c})^{m}\left\vert G\right\rangle $\ can be extended to $2N$,
provided that the extra conditions $(s_{a}+s_{c})^{N+m}\left\vert
G\right\rangle =0$\ and $(s_{b}+s_{c})^{N+m}\left\vert G\right\rangle =0$\
are added. Fourth, this conclusion can be extended to the case with multiple
sub-Hamiltonians. In the following, three illustrative examples are given to
demonstrate the theorem.

(i) Free-boson models on odd-sized chains. We consider two noninteracting
bosonic systems on chains with an odd number of sites. The corresponding
Hamiltonians for chains with $\left( 2N_{1}+1\right) $ sites and $\left(
2N_{2}+1\right) $ sites are given by
\begin{equation}
H_{1}=-\sum_{l=1}^{2N_{1}}(b_{l}^{\dagger }b_{l+1}+\mathrm{H.c.}),
\end{equation}%
and
\begin{equation}
H_{2}=-\sum_{l=2N_{1}+1}^{2N_{1}+2N_{2}}(b_{l}^{\dagger }b_{l+1}+\mathrm{H.c.%
}),
\end{equation}%
respectively, where $b_{l}$ is a boson operator satisfying the commutation
relations%
\begin{equation}
\left[ b_{j},b_{l}^{\dagger }\right] =\delta _{jl},\left[ b_{j},b_{l}\right]
=0.
\end{equation}%
Straightforward derivations show that the following relations hold
\begin{equation}
H_{1}\left[ \sum_{l=1}^{2N_{1}+1}\sin \left( \frac{l\pi }{2}\right)
b_{l}^{\dagger }\right] ^{m}\left\vert \text{\textrm{Vac}}\right\rangle =0,
\end{equation}%
and%
\begin{equation}
H_{2}\left[ \sum_{l=2N_{1}+1}^{2N_{1}+2N_{2}+1}\sin \left( \frac{l\pi }{2}%
\right) b_{l}^{\dagger }\right] ^{m}\left\vert \text{\textrm{Vac}}%
\right\rangle =0,
\end{equation}%
which respectively provide the zero eigenstates of $H_{1}$\ and $H_{2}$.
This indicates that the operators are%
\begin{eqnarray}
s_{a} &=&\sum_{l=1}^{2N_{1}}\sin \left( \frac{l\pi }{2}\right)
b_{l}^{\dagger },s_{b}=\sum_{l=2N_{1}+2}^{2N_{1}+2N_{2}+1}\sin \left( \frac{%
l\pi }{2}\right) b_{l}^{\dagger },  \notag \\
s_{c} &=&\sin \frac{(2N_{1}+1)\pi }{2}b_{2N_{1}+1}^{\dagger },
\end{eqnarray}
and\ $\left\vert G\right\rangle =\left\vert \text{\textrm{Vac}}%
\right\rangle $ is the vacuum state of the boson in this situation. On the
other hand, the zero-energy eigenstates of the Hamiltonian $H_{1}+H_{2}$ are
provided by the relation\ 
\begin{equation}
\left( H_{1}+H_{2}\right) \left[ \sum_{l=1}^{2N_{1}+2N_{2}+1}\sin \left( 
\frac{l\pi }{2}\right) b_{l}^{\dagger }\right] ^{m}\left\vert \text{\textrm{%
Vac}}\right\rangle =0,
\end{equation}%
which accords with the theorem.

(ii) Free-boson chains with ending potentials. Now we turn to consider two
noninteracting bosonic systems on chains with match ending potentials. The
corresponding Hamiltonians for chains with $N_{1}$ sites and $N_{2}+1$ sites
are given by
\begin{equation}
H_{1}=-\sum_{l=1}^{N_{1}-1}(b_{l}^{\dagger }b_{l+1}+\mathrm{H.c.}%
-e^{q_{1}}n_{l}^{b}-e^{-q_{1}}n_{l+1}^{b}),
\end{equation}%
and%
\begin{equation}
H_{2}=-\sum_{l=N_{1}}^{N_{1}+N_{2}-1}(b_{l}^{\dagger }b_{l+1}+\mathrm{H.c.}%
-e^{q_{2}}n_{l}^{b}-e^{-q_{2}}n_{l+1}^{b}),
\end{equation}%
respectively, where $n_{l}^{b}=b_{l}^{\dagger }b_{l}$ is boson number
operator. Straightforward derivations show that the following relations hold
\begin{equation}
H_{1}\left( \sum_{l=1}^{N_{1}}e^{q_{1}l}b_{l}^{\dagger }\right)
^{m}\left\vert \text{\textrm{Vac}}\right\rangle =0,
\end{equation}%
and%
\begin{equation}
H_{2}\left( \sum_{l=N_{1}}^{N_{1}+N_{2}}e^{q_{2}l}b_{l}^{\dagger }\right)
^{m}\left\vert \text{\textrm{Vac}}\right\rangle =0,
\end{equation}%
which respectively provide the zero eigenstates of $H_{1}$\ and $H_{2}$.
This indicates that the operators are
\begin{eqnarray}
s_{a} &=&\sum_{l=1}^{N_{1}-1}e^{q_{1}l}b_{l}^{\dagger
},s_{b}=\sum_{l=N_{1}+1}^{N_{1}+N_{2}}e^{q_{2}l}b_{l}^{\dagger },  \notag \\
s_{c} &=&e^{q_{1}N_{1}}b_{N_{1}}^{\dagger },
\end{eqnarray}%
under the constraint that
\begin{equation}
q_{1}=q_{2},
\end{equation}%
On the other hand, the zero-energy eigenstates of the Hamiltonian $%
H_{1}+H_{2}$ are provided by the relation\ we have%
\begin{equation}
\left( H_{1}+H_{2}\right) \left(
\sum_{l=1}^{N_{1}+N_{2}}e^{q_{1}l}b_{l}^{\dagger }\right) ^{m}\left\vert 
\text{\textrm{Vac}}\right\rangle =0,
\end{equation}%
which accords with the theorem. This provides a method to construct the
eigenstates of a large-size system using the eigenstates of several dimers.
In Fig. \ref{fig1}, several representative configurations for the
construction processes are schematically illustrated. Obviously, the
constructed eigenstates of $H_{1}+H_{2}$\ in both above examples are boson
condensate states, possesses ODLRO in thermodynamic limit, according to ref. 
\cite{Yang1962}. 

(iii) $\eta $-pairing states in Hubbard chains. The Hubbard model is a
fundamental model in condensed matter physics used to describe the behavior
of correlated electrons in a lattice. It is particularly useful for studying
phenomena such as superconductivity and magnetism. $\eta $-pairing states
are a special type of eigenstate of the Hubbard model that exhibit
off-diagonal long-range order (ODLRO) \cite{Yang1989}. These states are
characterized by the presence of Cooper-like pairs of electrons with
opposite spins, which is a key feature of superconductivity. The $\eta $%
-pairing states are particularly interesting because they can lead to
superconductivity under certain conditions. We consider two Hubbard models
on chains with $N_{1}$ sites and $N_{2}$ sites. The corresponding
Hamiltonians are given by
\begin{eqnarray}
H_{1} &=&-\sum_{l=1}^{N_{1}-1}\sum_{\sigma =\uparrow ,\downarrow }\left(
c_{l,\sigma }^{\dagger }c_{l+1,\sigma }+\text{\textrm{H.c.}}\right)  \notag
\\
&&+U\sum_{l=1}^{N_{1}}\left( n_{l,\uparrow }-\frac{1}{2}\right) \left(
n_{l,\downarrow }-\frac{1}{2}\right)  \notag \\
&&-\frac{1}{2}\sum_{\sigma =\uparrow ,\downarrow }\left( c_{N_{1},\sigma
}^{\dagger }c_{N_{1}+1,\sigma }+\text{\textrm{H.c.}}\right) ,
\end{eqnarray}%
and%
\begin{eqnarray}
H_{2} &=&-\sum_{l=N_{1}+1}^{N_{1}+N_{2}-1}\sum_{\sigma =\uparrow ,\downarrow
}\left( c_{l,\sigma }^{\dagger }c_{l+1,\sigma }+\text{\textrm{H.c.}}\right) 
\notag \\
&&+U\sum_{l=N_{1}+1}^{N_{1}+N_{2}}\left( n_{l,\uparrow }-\frac{1}{2}\right)
\left( n_{l,\downarrow }-\frac{1}{2}\right)  \notag \\
&&-\frac{1}{2}\sum_{\sigma =\uparrow ,\downarrow }\left( c_{N_{1},\sigma
}^{\dagger }c_{N_{1}+1,\sigma }+\text{\textrm{H.c.}}\right) .
\end{eqnarray}%
Here, $c_{i,\sigma }$ is the annihilation operator for an electron at site $%
i $ with spin $\sigma $, and $n_{i,\sigma }=c_{i,\sigma }^{\dagger
}c_{i,\sigma }$. The first two terms in the Hamiltonians $H_{1}$ and $H_{2}$
represent standard Hubbard models, while the third terms represent the
hopping term between site $N_{1}$\ and site $N_{1}+1$. The parameter $U$
represents the interaction energy scale in the unit of 1.

Introducing a set of $\eta $-operators, given by 
\begin{equation}
\eta _{j}=\left( -1\right) ^{j}c_{j,\uparrow }^{\dagger }c_{j,\downarrow
}^{\dagger },  \label{eta1}
\end{equation}%
we obtaine the relations%
\begin{equation}
H_{1}\left( \sum_{l=1}^{N_{1}+1}\eta _{j}\right) ^{m}\left\vert \text{%
\textrm{Vac}}\right\rangle =0,
\end{equation}%
and%
\begin{equation}
H_{2}\left( \sum_{l=N_{1}}^{N_{1}+N_{2}}\eta _{j}\right) ^{m}\left\vert 
\text{\textrm{Vac}}\right\rangle =0,
\end{equation}%
which demonstrate the existence of $\eta $-pairing\ eigenstates for $H_{1}$
and $H_{2}$. This indicates that the operators are defined as%
\begin{equation}
s_{a}=\sum_{l=1}^{N_{1}-1}\eta _{j},s_{b}=\sum_{l=N_{1}+2}^{N_{1}+N_{2}}\eta
_{j},s_{c}=\eta _{N_{1}}+\eta _{N_{1}+1}.
\end{equation}%
On the other hand, the $\eta $-pairing\ eigenstates of the Hamiltonian $%
H_{1}+H_{2}$ are provided by the relation%
\begin{equation}
\left( H_{1}+H_{2}\right) \left( \sum_{l=1}^{N_{1}+N_{2}}\eta _{j}\right)
^{m}\left\vert \text{\textrm{Vac}}\right\rangle =0,
\end{equation}
which accords with the theorem. The local Cooper-like pairs act as
hardcore bosons, which are the primary focus of this work. In addition, we
note that we have the commutation relations
\begin{equation}
\left[ H_{1}+H_{2},s_{a}+s_{b}+s_{c}\right] =0,
\end{equation}%
for the three examples, which seem to be a little trivial and are not
required in the theorem. In the following sections, we will focus on the
Hamiltonians that do not obey the extra symmetries arising from the above
commutation relations.

\section{Hardcore boson systems}

\label{Hardcore boson systems}

In this section, we focus on the hardcore boson Hamiltonian on a chain with
NN interaction. The investigation of this model is based on the theorem
proposed above. To this end, we start with the minimal-sized systems, based
on which the conclusions for the large-sized systems can be obtained. The
corresponding Hamiltonians for two hardcore boson dimers are given by
\begin{eqnarray}
H_{1} &=&-[a_{1}^{\dagger }a_{2}+\mathrm{H.c.}%
+(e^{-q_{1}}n_{1}+e^{q_{1}}n_{2})  \notag \\
&&+2\cosh q_{1}(n_{1}n_{2}-n_{1}-n_{2})],
\end{eqnarray}%
and%
\begin{eqnarray}
H_{2} &=&-[a_{2}^{\dagger }a_{3}+\mathrm{H.c.}%
+(e^{-q_{2}}n_{2}+e^{q_{2}}n_{3})  \notag \\
&&+2\cosh q_{2}(n_{2}n_{3}-n_{2}-n_{3})],
\end{eqnarray}%
respectively, where $a_{l}^{\dagger }$ is the hardcore-boson creation
operator at the position $l$, and $n_{l}=a_{l}^{\dagger }a_{l}$. The
hardcore-boson operators satisfy the commutation relations%
\begin{equation}
\left\{ a_{l},a_{l}^{\dagger }\right\} =1,\left\{ a_{l},a_{l}\right\} =0,
\end{equation}%
and%
\begin{equation}
\left[ a_{j},a_{l}^{\dagger }\right] =0,\left[ a_{j},a_{l}\right] =0,
\end{equation}%
for $j\neq l$. The total particle number operator, $n=\sum_{l}n_{l}$, is a
conserved quantity because it commutes with the Hamiltonian. Then one can
investigate the system in each invariant subspace with fixed particle number 
$n$. The model can be mapped to the spin-$1/2$\ 
XXZ model 
\cite{Matsubara1956Alatticemodel}, 
which enables the application of our results to both hard-core
boson and quantum spin systems. 

Straightforward derivations show that the following relations hold
\begin{equation}
H_{1}\left( e^{q_{1}}a_{1}^{\dagger }+e^{2q_{1}}a_{2}^{\dagger }\right)
^{m}\left\vert \text{\textrm{Vac}}\right\rangle =0,
\end{equation}%
and%
\begin{equation}
H_{2}\left( e^{2q_{2}}a_{2}^{\dagger }+e^{3q_{2}}a_{3}^{\dagger }\right)
^{m}\left\vert \text{\textrm{Vac}}\right\rangle =0,
\end{equation}%
which respectively provide the zero eigenstates of $H_{1}$\ and $H_{2}$.
Here, there is no restriction on the integer $m$. In fact, we always have
the identities $(e^{q_{1}}a_{1}^{\dagger }$ $+e^{2q_{1}}a_{2}^{\dagger
})^{m} $ $\left\vert \text{\textrm{Vac}}\right\rangle =0$\ and $%
(e^{2q_{2}}a_{2}^{\dagger }$ $+e^{3q_{2}}a_{3}^{\dagger })^{m}$ $\left\vert 
\text{\textrm{Vac}}\right\rangle =0$\ for $m>1$. These results allow us to
construct two sets of operators%
\begin{equation}
s_{a}=e^{q}a_{1}^{\dagger },s_{b}=e^{3q}a_{3}^{\dagger
},s_{c}=e^{2q}a_{2}^{\dagger },
\end{equation}%
by taking $q_{1}=q_{2}=q$, and%
\begin{equation}
s_{a}=e^{q}a_{1}^{\dagger },s_{b}=e^{q}a_{3}^{\dagger
},s_{c}=e^{2q}a_{2}^{\dagger },
\end{equation}%
by taking $q_{1}=-q_{2}=q$,\ respectively. According to the theorem, two
sets of zero-energy eigenstates of the Hamiltonian $H_{1}+H_{2}$ can be
provided by the relations\ 
\begin{equation}
\left( H_{1}+H_{2}\right) (e^{q}a_{1}^{\dagger }+e^{2q}a_{2}^{\dagger
}+e^{3q}a_{3}^{\dagger })^{m}\left\vert \text{\textrm{Vac}}\right\rangle =0,
\end{equation}%
and%
\begin{equation}
\left( H_{1}+H_{2}\right) (e^{q}a_{1}^{\dagger }+e^{2q}a_{2}^{\dagger
}+e^{q}a_{3}^{\dagger })^{m}\left\vert \text{\textrm{Vac}}\right\rangle =0,
\end{equation}%
respectively. This also provides a method to construct the eigenstates of a
large-size interacting system using the eigenstates of several dimers. In
Fig. \ref{fig1}, several representative configurations for the construction
processes are schematically illustrated. This indicates that one can
construct two types of trimers based on those of two dimers, which possess a
set of zero-energy eigenstates. In addition, we note that
\begin{eqnarray}
&&\left[ H_{1},s_{a}+s_{c}\right] \neq 0,\left[ H_{2},s_{b}+s_{c}\right]
\neq 0,  \notag \\
&&\left[ H_{1}+H_{2},s_{a}+s_{b}+s_{c}\right] \neq 0,
\end{eqnarray}%
that is, such a construction does not require the extra commutation
relations. This is crucial for the study of quantum many-body scars. Quantum
many-body scar states are many-body states with finite energy density in
non-integrable models that do not obey the eigenstate thermalization
hypothesis. A tower of scar states is equally spaced in energy \cite%
{vosk2013many}. It has been pointed out that such a tower can be constructed
using the restricted spectrum generating algebra, which is not based on
symmetry \cite{moudgalya2020eta}. Although the constructed eigenstates are
degenerate, they will form an energy tower when a uniform chemical potential
is added.

\section{Condensate ground states with ODLRO}

\label{Condensate ground states with ODLRO}

In this section, as applications, we will construct Hamiltonians on
large-sized lattices using the two types of trimers as building blocks. We
focus on two types of large-sized systems. One is a uniform chain embedded
with a single impurity. The other is a periodic chain with an impurity array.

\subsection{Local condensate ground states}

\label{Local condensate ground states}

We consider the Hamiltonian in the form
\begin{eqnarray}
&&H_{\text{\textrm{sng}}}=-\sum_{l=-(N-1)}^{N}(a_{l-1}^{\dagger}a_{l}+\mathrm{H.c.}+2\cosh{q}n_{l-1}n_{l})  \notag \\
&&-[2\sinh qn_{0}+e^{-q}\left( n_{N}+n_{-N}\right) 
\notag \\ &&-2\cosh
q\sum_{l=-N}^{N}n_{l}],
\end{eqnarray}%
which describes a uniform chain of length $2N+1$ with three on-site
potentials at the center and two ends, in addition to a global on-site
potential. It is schematically illustrated in Fig. \ref{fig1}(d). It has
been shown that the ending potentials $e^{-q}$\ ($q>0$) cannot form bound
states for large $N$ and can therefore be neglected in the large $N$ limit.
In fact, a set of eigenstates can be obtained by applying the proposed
theorem to this Hamiltonian.

To this end, we rewrite the Hamiltonian in a specific form, that is,%
\begin{equation}
H_{\text{\textrm{sng}}}=\sum_{l=1}^{N}\left( H_{l}^{-}+H_{l}^{+}\right) ,
\end{equation}%
as the summation of dimers, where 
\begin{eqnarray}
&&H_{l}^{\pm }=-a_{\pm \left( l-1\right) }^{\dagger }a_{\pm l}+\mathrm{H.c.}%
+e^{q}n_{\pm \left( l-1\right) }+e^{-q}n_{\pm l}  \notag \\
&&+2\cosh q\left( n_{\pm \left( l-1\right) }n_{\pm l}-n_{\pm \left(
l-1\right) }-n_{\pm l}\right) .
\end{eqnarray}%
According to the above analysis, the ground state in $m$-particle invariant
subspace can be obtained as%
\begin{equation}
\left\vert \psi _{\text{\textrm{g}}}^{m}\right\rangle =\left( \sum_{l=-N}^{N}a_{l}^{\dagger }e^{-\lvert l\rvert q}\right)
^{m}\left\vert \text{\textrm{Vac}}\right\rangle ,  \label{local}
\end{equation}%
with zero groundstate energy.

In order to gain insight into the features of the state $\left\vert \psi _{%
\text{\textrm{g}}}^{m}\right\rangle $, we consider the case with larger $N$
and $q$. The state $\left\vert \psi _{\text{\textrm{g}}}^{m}\right\rangle $\
can be approximately written in the following form%
\begin{eqnarray}
&&\left\vert \psi _{\text{\textrm{g}}}^{1}\right\rangle \approx
a_{0}^{\dagger }\left\vert \text{\textrm{Vac}}\right\rangle ,  \notag \\
&&\left\vert \psi _{\text{\textrm{g}}}^{2}\right\rangle \approx
(a_{0}^{\dagger }a_{1}^{\dagger }+a_{0}^{\dagger }a_{-1}^{\dagger
})\left\vert \text{\textrm{Vac}}\right\rangle ,  \notag \\
&&\left\vert \psi _{\text{\textrm{g}}}^{3}\right\rangle \approx
a_{-1}^{\dagger }a_{0}^{\dagger }a_{1}^{\dagger }\left\vert \text{\textrm{Vac%
}}\right\rangle ,  \notag \\
&&\vdots  \notag \\
&&\left\vert \psi _{\text{\textrm{g}}}^{2k+1}\right\rangle \approx
a_{0}^{\dagger }\prod_{l=1}^{k}a_{-l}^{\dagger }a_{l}^{\dagger }\left\vert 
\text{\textrm{Vac}}\right\rangle ,  \notag \\
&&\left\vert \psi _{\text{\textrm{g}}}^{2k+2}\right\rangle \approx \left(
a_{-k-1}^{\dagger }+a_{k+1}^{\dagger }\right) a_{0}^{\dagger
}\prod_{l=1}^{k}a_{-l}^{\dagger }a_{l}^{\dagger }\left\vert \text{\textrm{Vac%
}}\right\rangle .
\end{eqnarray}%
Obviously, the state $\left\vert \psi _{\text{\textrm{g}}}^{m}\right\rangle $
represents a state with $m$ particles frozen around the central impurity,
acting as an insulating domain. The size of the domain equals the particle
number $m$. Specifically, for a finite $N$, we always have $\left\vert \psi
_{\text{\textrm{g}}}^{2N+1}\right\rangle =a_{0}^{\dagger
}\prod_{l=1}^{N}a_{-l}^{\dagger }a_{l}^{\dagger }\left\vert \text{\textrm{Vac%
}}\right\rangle $ for any given value of $q$. On the other hand, considering
the case with small $m$ and $q$, there exist quantum fluctuations around the
insulating domain in the ground state. Such quantum fluctuations play an
important role in the case with multiple impurities.

\subsection{Extended condensate ground states}

\label{Extended condensate ground states}

In this section, we focus on constructing periodic Hamiltonians based on the
above result. These Hamiltonians are designed to allow the system to possess
extended condensate ground states, thereby supporting ODLRO. The underlying
mechanism is simple. If we consider a Hamiltonian constructed from the
summation of two distinct sets of dimers, its structure can be periodic.
Then the corresponding condensate eigenstates are extended states.

Specifically, consider the Hamiltonian on a $2N$-site ring, given by
\begin{eqnarray}
H_{\text{\textrm{arr}}} &=&-\sum_{l=1}^{2N}[a_{l}^{\dagger }a_{l+1}+\mathrm{%
H.c.}+2\cosh qn_{l}n_{l+1}]  \notag \\
&&+2\sum_{l=1}^{N}\left( e^{q}n_{2l-1}+e^{-q}n_{2l}\right) ,
\end{eqnarray}
where the periodic boundary condition $a_{2N+1}=a_{1}$\ is taken. The
Hamiltonian describes a hardcore-boson model with NN interaction and
staggered on-site potentials in addition to a global on-site potential. It
is schematically illustrated in Fig. \ref{fig1}(e). Apparently, it is
difficult to obtain the solutions of the Hamiltonian except in the
single-particle invariant subspace. However, we will show that the ground
state in the m-particle invariant subspace can be constructed from the one
in the single-particle invariant subspace.

To this end, we rewrite the Hamiltonian in a specific form, that is, 
\begin{equation}
H_{\text{\textrm{arr}}}=\sum_{l=1}^{N}\left( H_{l}^{-}+H_{l}^{+}\right) ,
\end{equation}%
as the summation of dimers, where%
\begin{eqnarray}
H_{l}^{\pm } &=&-[a_{2l}^{\dagger }a_{2l\pm 1}+\mathrm{H.c.}%
+e^{q}n_{2l}+e^{-q}n_{2l\pm 1}  \notag \\
&&+2\cosh q\left( n_{2l}n_{2l\pm 1}-n_{2l}-n_{2l\pm 1}\right) ].
\end{eqnarray}%
According to the above analysis, the ground state in $m$-particle invariant
subspace can be obtained as%
\begin{equation}
\left\vert \psi _{\text{\textrm{g}}}^{m}\right\rangle =\left[
\sum_{l=1}^{N}(e^{q}a_{2l}^{\dagger }+a_{2l-1}^{\dagger })\right]
^{m}\left\vert \text{\textrm{Vac}}\right\rangle ,  \label{extended}
\end{equation}%
with zero groundstate energy.

In order to gain insight into the features of the state $\left\vert \psi _{%
\text{\textrm{g}}}^{m}\right\rangle $, we consider the case with larger $q$.
The state $\left\vert \psi _{\text{\textrm{g}}}^{m}\right\rangle $\ can be
approximately written in the following form
\begin{equation}
\left\vert \psi _{\text{\textrm{g}}}^{m}\right\rangle \approx \left(
\sum_{l=1}^{N}a_{2l}^{\dagger }\right) ^{m}\left\vert \text{\textrm{Vac}}%
\right\rangle ,
\end{equation}%
where an overall factor is neglected. The correlation function, defined as%
\begin{equation}
C(l,l^{\prime })=\left\langle \phi \right\vert a_{l}^{\dagger }a_{l^{\prime
}}\left\vert \phi \right\rangle ,
\end{equation}%
can be introduced to measure the nature of condensation for a given state $%
\left\vert \phi \right\rangle $. In the large $N$ limit, the correlation
function for the ground state $\left\vert \psi _{\text{\textrm{g}}%
}^{m}\right\rangle $\ is estimated as%
\begin{eqnarray}
C_{\text{\textrm{g}}}^{m}(2l,2l+2r) &=&\frac{\left\langle \psi _{\text{%
\textrm{g}}}^{m}\right\vert a_{2l}^{\dagger }a_{2l+2r}\left\vert \psi _{%
\text{\textrm{g}}}^{m}\right\rangle }{\left\langle \psi _{\text{\textrm{g}}%
}^{m}\right\vert \left. \psi _{\text{\textrm{g}}}^{m}\right\rangle }  \notag
\\
&\approx &\frac{\left( N-m\right) m}{N^{2}}.
\end{eqnarray}%
It indicates that state $\left\vert \psi _{\text{\textrm{g}}%
}^{m}\right\rangle $\ possesses ODLRO according to ref. \cite%
{Yang1962} due to the fact that the correlation function does not
decay as $r$\ increases for finite particle density $m/N$. It is presumably
the case that when a smaller $q$ is considered, the fluctuations increase in
the ground state. However, it cannot affect too much the nature of ODLRO.

This method and conclusion can be extended to other superlattice structures,
which consist of a multiple-site unit cell. For instance, a $4$-site unit
cell may contain $2$ impurities. Such a supperlattice system can be
constructed by the building block given in Fig. \ref{fig1}(b2).

\section{Dynamic stability}

\label{Dynamic stability}

It has been shown that the conclusion in the last section depends on the
resonance between the on-site potentials and the strength of the NN
interactions in these systems. The condensate state $\left\vert \psi _{\text{%
\textrm{g}}}^{m}\right\rangle $\ is the exact ground state of the hardcore
bose-Hubbard Hamiltonian. However, the off-resonance on-site potential
strength may lead to a deviation of the eigenstate from the expression of $%
\left\vert \psi _{\text{\textrm{g}}}^{m}\right\rangle $, and this deviation
is usually considered in practice. In this section, we focus on the
influence of the off-resonance effect on the existence of the long-range
order condensate state.

\begin{figure*}[tbph]
\centering
\includegraphics[width=1.0\textwidth]{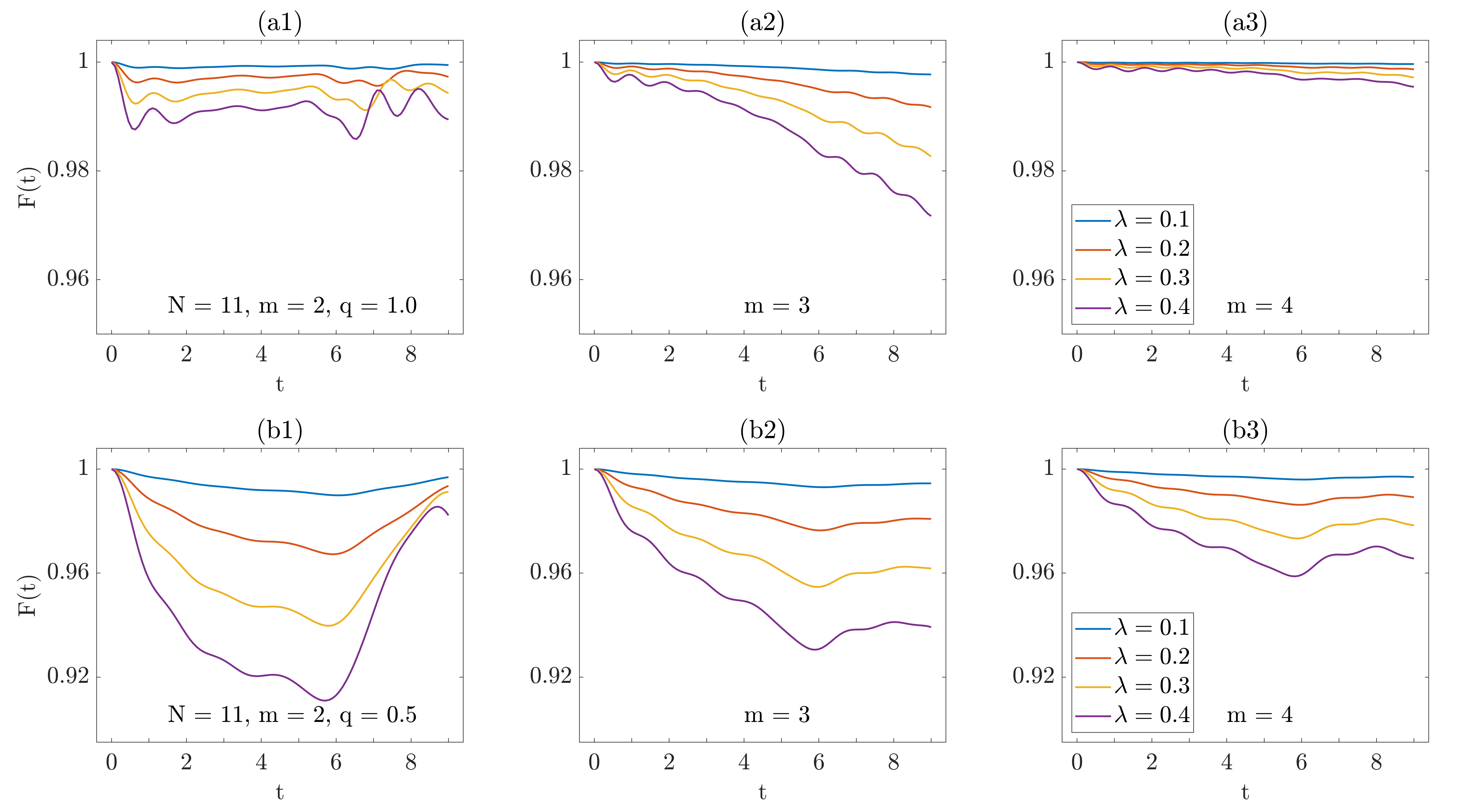}
\caption{Plots of the fidelity $F(t)$ defined in Eq. (\protect\ref{fidelity}%
) for the Bose-Hubbard chains with (a1)-(a3) $q=1$\ and (b1)-(b3) $q=0.5$,
respectively. The results are obtained by exact diagonalization for
finite-size system with $N=11$. The system parameters $m$\ and $\protect%
\lambda $ are indicated in the panels. As expected, the fidelity $F(t)$
decays with time $t$ for finite values of $\protect\lambda $. The results
indicate that the fidelity $F(t)$ decays less rapidly for small values of $%
\protect\lambda $, especially for larger values of\ $q$ and $m$.}
\label{fig2}
\end{figure*}

Our strategy is to examine the dynamic response of the ground state $%
\left\vert \psi _{\text{\textrm{g}}}^{m}\right\rangle $ under a quenching
process. Specifically, we numerically compute the time evolution of the
state $\left\vert \phi \left( 0\right) \right\rangle =\left\vert \psi _{%
\text{\textrm{g}}}^{m}\right\rangle $\ as the initial state under the quench
Hamiltonian in the form%
\begin{equation}
H_{\text{\textrm{qnc}}}=H_{\text{\textrm{sng}}}+\lambda e^{q}n_{0},  
\end{equation}%
where $\lambda $\ characterizes the the strength the off-resonance.\ The
evolved state can be expressed as%
\begin{equation}
\left\vert \phi \left( t\right) \right\rangle =e^{-iH_{\text{\textrm{qnc}}%
}t}\left\vert \psi _{\text{\textrm{g}}}^{m}\right\rangle ,
\end{equation}%
which is calculated by exact diagonalization for finite systems\ with
several typical particle filling numbers $m$. We employ the fidelity, given by
\begin{equation}
F(t)=\left\vert \left\langle \psi _{\text{\textrm{g}}}^{m}\right\vert
e^{-iH_{\text{\textrm{qnc}}}t}\left\vert \psi _{\text{\textrm{g}}%
}^{m}\right\rangle \right\vert ^{2},  \label{fidelity}
\end{equation}%
to characterize the dynamic response\ induced by different values of $%
\lambda $. In this work, we only focus on the ground state $\left\vert \psi
_{\text{\textrm{g}}}^{m}\right\rangle $\ in Eq. (\ref{local}), and the
results obtained can also shed light on those of the periodic system. {We
plot }$F(t)${\ in Fig. \ref{fig2} as a function of }${t}${\ for selected
systems and particle numbers. The results show }that the fidelity $F(t)$
decays with time $t$ for finite values of $\lambda $ as expected, indicating
the necessity of the resonance. However, we observe that the fidelity $F(t)$
decays less rapidly for larger values of\ $q$, indicating the stability of
the ground states in cases of near resonance. In addition, for fixed values
of $\lambda $\ and $q$, the ground states become more stable as $n$
increases.

\section{Summary}

\label{Summary}

In summary, we have proposed a general method to construct the condensate
eigenstates from those of sub-Hamiltonians. This method is applicable to
many-body systems, including interacting fermionic, bosonic, and quantum
spin systems. Importantly, it provides the possibility to construct the
Hamiltonian possessing condensate ground states. In this sense, it
performs a similar task as real-space renormalization group methods 
\cite{kadanoff1966scaling,wilson1975renormalization,white1992density}. We exemplified this finding through the investigation of a concrete system:
an extended hardcore Bose-Hubbard model on one-dimensional lattices. We
demonstrated that a local on-site potential can counteract the hardcore
effect and induce an evanescent condensate mode. Based on this, we
constructed a superlattice system by applying an array of impurities. The
exact condensate ground states of hardcore bosons with a fixed boson number
were obtained and shown to possess ODLRO. This conclusion can be extended to
higher-dimensional systems. Additionally, we investigated the effect of the
off-resonance strength of the on-site potentials on the condensate ground
states using numerical simulations of the dynamic response. Our results
provide an alternative way to explore novel systems with condensate phases.

\section*{ACKNOWLEDGMENTS}

This paper was supported by the National Natural Science Foundation of China (under Grant No. 12374461).


\end{document}